\documentclass[conference]{IEEEtran}
\usepackage{cite}
\usepackage{amsmath,amssymb,amsfonts}
\usepackage{graphicx}
\usepackage{textcomp}
\usepackage{multicol}
\usepackage{multirow}
\usepackage{subfigure}
\usepackage{float}
\usepackage{booktabs}
\usepackage[colorlinks=true, allcolors=blue]{hyperref}

\title{Techreport: Evaluating Tor-based Location Privacy for Ethereum Validators}

\author{Muhammad Umar Janjua$^1$, Akshaya Mani$^1$, Uğur Şen$^{*1}$, Daniel Kaiser$^1$ \\ 
$^1$ Institute of Free Technology \\ 
 umarjanjua@live.com \\ 
\{akshaya,ugur,ksr\}@status.im

}

\begin{document}
\maketitle

\begin{abstract}
Privacy and anonymity of validators, especially regarding IP address linkability, are essential to protect the Ethereum network from various attacks. Network-level attacks, such as DoS, can interrupt validators and affect the overall security of the Ethereum network. Correlating the IP addresses of validators with their identities, along with knowledge about their action slots can be exploited by attackers to cause network delays, MEV exploitation, and finality risks. Therefore, ensuring the unlinkability of a validator's IP and identity is crucial for maintaining the network's trust and resilience. In this techreport, we first provide a review of the existing network and consensus layer techniques that have been proposed for maintaining validator privacy in the Ethereum blockchain. Secondly, we evaluate a Tor-based protocol named \emph{Tor push} that helps unlink validator identities (IDs) from their nodes' IP addresses, thereby making it difficult to determine any end-to-end correlation between validator IDs and IP addresses of validators' beacon nodes. To evaluate the effectiveness of \emph{Tor push}, we present a working, deployed proof-of-concept (PoC) implementation in the Nimbus Ethereum client. Our PoC deployment pushes attestations, aggregations, and block proposals over Tor to the Goerli testnet. Furthermore, we also analyse the security and latency of \emph{Tor push}. Our experimental results suggest that Tor can be incorporated into the existing Ethereum network with a tolerable latency overhead of 613.82 ms on average and without compromising the overall network performance while enhancing the location privacy of validators in the Ethereum network.
\end{abstract}

\section{Introduction}
The privacy and anonymity of validators in the Ethereum network are important for the overall security, integrity, and decentralization of the blockchain ecosystem. Attackers can be motivated to correlate IP addresses with validators' IDs and pinpoint the timing of validators' actions for a range of malicious purposes. These attacks encompass a broader landscape of network-level threats and can significantly impact the Ethereum network's stability and user trust. For example, maintaining the location privacy of validators could be helpful against Sybil attacks \cite{9236965}, where malicious actors strategically position multiple nodes in proximity to validators. Eclipse attacks \cite{190890} and other network-level threats, involve isolating a target node, such as a validator, from the rest of the network. Attackers equipped with knowledge about a validator's location can manipulate network routing to isolate the validator. Once isolated, they can execute Distributed Denial of Service (DDoS) attacks, potentially causing delays in validator duties/operations, which may affect the chain's finality and integrity.

Validator sniping is another scenario where a participant node tries to strategically time its actions to become a validator at a particular slot, either to maximize rewards or to influence the network in some way \cite{Brgel_2023}. This becomes a threat when the location and actions of validators are exposed. Malicious actors can target validators with substantial stakes, coercing them to miss blocks or act in ways that undermine the network. Furthermore, attackers may resort to DDoS attacks when they know the validator's IP address. These attacks can disrupt a validator's operation, cause delay, or even force it offline, compromising network security and integrity. An attacker can also deploy DDoS attacks strategically to coerce a validator into missing blocks, leading to financial losses and adverse effects on finality and fairness. For example, if an attacker successfully links a validator ID to its corresponding IP address, then the attacker will be able to identify the proposer that has been selected for the next slot. In this way, the attacker could learn the IP address of the selected validator and then realize a DoS attack on the selected validator, making it impossible to propose the block in the selected slot.

Furthermore, the Miner Extractable Value (MEV) strategy that revolves around the profit that miners or validators can extract by manipulating the order of transactions can be affected. Knowing validators's IP addresses, if attackers disrupt or even delay validator activity, they can gain an advantage in conducting a front-running attack, placing their transactions in a way that maximizes their profit at the expense of others. This not only negatively affects the fairness of the system but can also erode trust in the network.

In the literature, various techniques have been proposed to maintain the anonymity of validators in the Ethereum network, which can be broadly categorized into two classes: (1) network layer techniques and (2) consensus layer techniques. Network layer techniques add an extra layer of anonymity and security to the communication between validators to prevent attackers from establishing a link between validators' IDs and their actions in the Ethereum network whereas consensus layer methods protect anonymity during the consensus process. This work builds upon the existing \emph{Tor push} \cite{vac46GOSSIPSUBTORPUSH} proposal by providing its first empirical deployment and evaluation in a live Ethereum test network. The objectives of this techreport are twofold: (1) providing an overview of existing techniques that have been used to ensure the unlinkability of validators' IDs and IP addresses, and (2) evaluating \emph{Tor push} which provides location privacy for Ethereum validators. Although there are many broader aspects of anonymity and privacy of validators, in this techreport, we specifically focus on ensuring the unlinkability of validators' IP addresses and their IDs. The following lists the main contributions of this techreport.

\begin{enumerate}
    \item We present a review of the existing techniques that have been proposed for enhancing validators' privacy in Ethereum primarily at the network-layer level. We briefly discuss consensus-level techniques for completeness.
    \item We provide a working, deployed, live proof-of-concept implementation of 10 Tor-based validators in Nimbus client that successfully pushes attestations, aggregations, and blocks on the Ethereum test network, Goerli with an average effectiveness score of 90\% (see Fig. \ref{fig:effectiveness}).
    \item We conduct a performance evaluation of the \emph{Tor push}  solution both theoretically and quantitatively. Theoretical analysis involves examining potential attack scenarios for security assessment, while quantitative evaluation measures the average latency overhead incurred by Tor.
\end{enumerate}

\textit{Organization of the Techreport:} The rest of the techreport is organized as follows. Section \ref{sec:back} presents the related work. Section \ref{sec:proposed} provides background on \emph{Tor push}. Experimental setup, security analysis, and results for performance evaluation of \emph{Tor push}  are discussed in Section \ref{sec:results}. Finally, we conclude the techreport in Section \ref{sec:concs}.

\section{{Related Work}}
\label{sec:back}
In this section, we provide an overview of existing techniques that can be used to ensure validators' privacy. We will also briefly discuss the trade-off between critical network parameters, i.e., anonymity, latency, and bandwidth.








\subsection{Existing Methods for Enhancing Validators Privacy}
\begin{figure}[!h]
    \centering
    \includegraphics[width=0.45\textwidth]{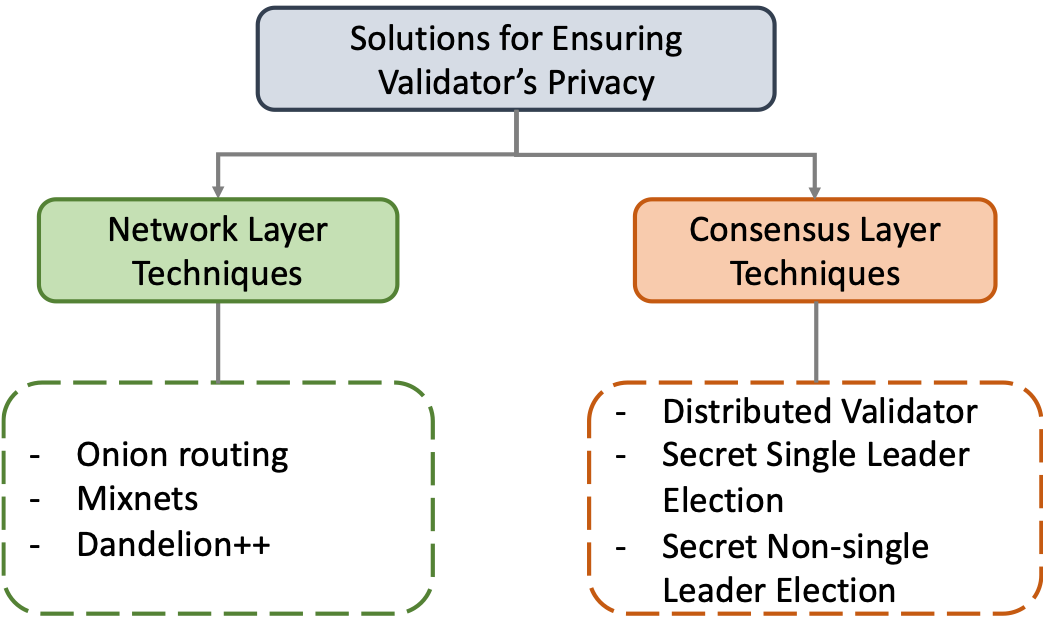}
    \caption{Different solutions for improving validators' privacy in the blockchain.}
    
    \label{fig:solutions_tax}
\end{figure}

In this section, we will discuss various existing solutions that have been proposed for improving validators' privacy in blockchain peer-to-peer (P2P) networks. In the literature, various studies have attempted to address the validators' privacy issues using different techniques. These techniques can be broadly classified into two categories, i.e., network-layer techniques and consensus-layer methods (as shown in Fig. \ref{fig:solutions_tax}). 
We start by first discussing network layer techniques.

\subsubsection{Network Layer Solutions for Validators' Privacy}
The major network layer solutions proposed for validators' privacy are (1) Tor/Onion Routing; (2) Mixnets; and (3) Dandelion/Dandelion++. 



\paragraph{Tor/Onion Routing}
Onion routing is a network layer technique, which is also used in the Tor network. It enables anonymous communication by allowing the sender and the receiver to establish a secure communication by passing the information through a series of layers (encryption) such that no one can confidently link the sender's identity with the recipient or to the message being communicated. In a recent study, Kelen et al. \cite{kelen2023integrated} suggested integrating onion routing to preserve the privacy of validators in the blockchain network and they investigated whether an onion routing-based method can be potentially used for making validator public keys (on-chain identities) unlinkable to their IP address (off-chain identities) in the Proof of Stake (PoS) consensus protocol in the Ethereum network. They discussed only theoretical guarantees of using Tor/Onion routing for validator privacy. 

The onion routing framework to preserve validators' privacy, proposed in \cite{kelen2023integrated}, works by first selecting at least three nodes using a uniform distribution from P2P network participants such that the information about the selected nodes is not revealed to any other participant. Then an encrypted communication channel is established between the originator and the first relayer, facilitating the exchange of shared secrets and public keys. After establishing this communication channel, the process is repeated with a second relayer (this is performed through an already established channel between the originator and the first relayer, which is relaying messages between the originator and the second relayer). When the channel between the second relayer and the originator has been established, this process is repeated for the third and fourth relayer, and it continues in the same fashion. Each time the communication channel established with the previous relayer is used such that the final channel between the originator and the broadcaster uses all relayers. The established channels are maintained by exchanging keepalive messages and spam-protection proofs. The established channels are used by the originator whenever it wants to propose a block or broadcast attestations. Kelen et al. \cite{kelen2023integrated} concluded that onion routing can potentially enhance validator privacy in the Ethereum blockchain. 

\paragraph{Mixnets}
Mixnets are routing techniques that provide privacy-preserving communication by ensuring that the link between the sender and receiver is obscured. Specifically, it uses a chain of proxies (i.e., intermediary nodes--often referred to as ``mix servers'' or ``mixes'') to route the message from sender to receiver making it difficult for adversaries to establish this relationship. Mixnets and Tor almost work similarly, however, there are a few key differences, e.g., mixnets have a mechanism to mitigate correlation attacks aiming to analyze network traffic. Mixnets can be potentially employed in the Ethereum network to enhance privacy, e.g., ensuring anonymity of validators, transaction privacy, and privacy of the whole network. Prominent implementations of Mixnets include NYM \cite{nym2023} and HOPR \cite{hopr2023}. HOPR network consists of a decentralized incentivized mixnet that ensures the privacy of data and metadata for its users. Relaying messages through multiple hops and data transfers is conducted in a manner that obscures the origin, destination, and content from anyone other than the sender and receiver. Similarly, ``NYM'' uses Mixnets to enhance privacy at the network layer. The following are the key properties of ``NYM'' protocol that enhance privacy:
\begin{enumerate}

    \item \textit{Multiple Hops:} Similar to Tor, messages are relayed through multiple hops to obfuscate the origin and destination. 

    \item \textit{Cover Traffic:} Cover (dummy) traffic is added to prevent traffic analysis, where all traffic data packets have the same size. 

    \item \textit{Time Obfuscation:} All data packets, including dummy (cover) packets, are mixed with those of other users. This mixing occurs at each hop, where packets take different routes, resulting in varying latencies at each stage.

    \item \textit{Horizontal Scalability:} Unlike blockchain networks, NYM Mixnet can be expanded according to the network needs, i.e., to allow more traffic (similar to the web). 
\end{enumerate}

\paragraph{Dandelion++} 
\noindent The Dandelion technique was developed for the privacy of cryptocurrency transactions on the Bitcoin P2P network \cite{bojja2017dandelion}. Its key objective is to make it harder to link a transaction with the IP address of the originating node, enhancing validators' privacy \cite{sharma2022anonymity}. Dandelion operates in two phases: In the anonymity (stem) phase, messages are relayed over proxies using a non-interactive graph creation technique. In the spreading (fluff) phase, messages are diffused over the main P2P network. Dandelion++ is an advanced version addressing idealistic assumptions of Dandelion \cite{fanti2018dandelion++}. These assumptions include: (1) each node generates one transaction, (2) strict adherence to the Dandelion protocol by all nodes, and (3) all nodes executing the Dandelion protocol. Violations of these assumptions can lead to deanonymization attacks. Similar to Dandelion, Dandelion++ has two phases, with differences in the anonymity phase, network propagation phase, and anonymity set. It extends Dandelion by incorporating capabilities like collaborative stem nodes and dynamic propagation duration to enhance transaction privacy. The efficacy of both methods depends on various factors such as network conditions and behavior.

Dandelion \cite{bojja2017dandelion} and Dandelion++ \cite{fanti2018dandelion++} work by differentiating the roles of the originator and the broadcaster, where the message from the originator arrives at the broadcaster after passing through a series of hops (which is a relayer that transmits the message to the next node) in the P2P network. However, in a subsequent study \cite{sharma2022anonymity}, the authors argued relayers are inadequate to prevent such attacks. Onion routing has some similarities to Dandelion, where the message from an originator to a broadcaster is relayed through several hops. However, the critical difference is that the encrypted message is relayed and only the broadcaster is capable of seeing the message in plain text. The originators' public key is used to encrypt a message for each relayer then the message is forwarded by each relayer after decryption using their private key.

\paragraph{Other Unexplored Techniques for Validator Privacy}

Some other anonymization networks could also be potentially used to enhance validator privacy such as I2P \cite{geti2p2023}. I2P is designed as an overlay network that allows anonymous communication between participants. While I2P has not been explored for protecting validator privacy in Ethereum, it offers a promising avenue. For example, Ethereum validators could route their network communications—including block proposals, attestations, and aggregations—through the I2P network to obfuscate their IP addresses. This could potentially mitigate the threat of adversaries attempting to identify and target validators based on their IP addresses, enhancing the overall security of the network. However, further research and experiments would be required to assess the feasibility and effectiveness of integrating I2P into Ethereum.

\subsubsection{Consensus Layer Solutions for Validators' Privacy}
Here, we will briefly discuss consensus layer solutions in the literature to enhance the privacy of Ethereum validators.

\paragraph{Distributed Validator (DV)} 
In this method, various tasks related to operating an Ethereum validator are spread across different multiple nodes while maintaining redundancy \cite{dls2023}. In this way, the DV method makes it significantly challenging for an attacker to realize a DoS attack on the validator that has been selected for proposing a block at a particular slot. Two key concepts are used in the DV technique: (1) consensus---validator's responsibilities are distributed among multiple co-validators that must work in collaboration to reach an agreement to sign any message (which is done through voting); and (2) M-of-N threshold signatures---staking key of the validator is divided into N parts and each share is assigned with the co-validators. When at least M of the co-validators reaches an agreement on a voting decision, each of them signs the message with their key share, and finally, a combined signature is reconstructed using the key shares held by them, however, DVT does not provide complete anonymity. 

\begin{table*}
\centering
\caption{Comparison of techniques for improving validators' privacy in the blockchain networks.}
\scalebox{0.95}{
\begin{tabular}{llp{6cm}p{6.5cm}}
\toprule
\multicolumn{1}{c}{\textbf{Operation}} & \multicolumn{1}{c}{\textbf{Technique}} & \multicolumn{1}{c}{\textbf{Description}} & \multicolumn{1}{c}{\textbf{Limitations}} \\
\midrule
\multirow{6}{*}{\textbf{Network Layer}} 
& Dandelion 
& Relays transactions through proxies: stem and fluff phases.
& - No unlinkability guarantees\\ 
  & Dandelion++ 
  & Forwards messages to a group of nodes. 
  & \parbox{6cm}{- Increased latency (due to stem phase)\\
  - No unlinkability guarantees. }\\
  & Tor/Onion Routing 
  & Relays messages through the Tor circuits or layers of encryptions. 
  & - Increased latency (due to encryption at multiple hops). \\
  & {Lightning Network} & {LN is a scalability solution proposed for Bitcoin, facilitating off-chain payment channels between entities with minimized fees.} & {- Limited security and transaction anonymity.} \\
  & {Mixnets} & {Employs intermediary nodes or "mix servers" to obscure the link between senders and receivers.} & {- Ensuring scalability and efficiency is challenging in blockchain networks.} \\
\midrule

\multirow{3}{*}{\textbf{Consensus Layer}} 
& Distributed Validators & Validator tasks are decentralized among multiple co-validators to enhance redundancy and reduce Denial of Service (DoS) risks. & \parbox{6cm}{- Coordination complexity among co-validators. \\ - Scalability can be challenging. \\ - May increase performance overhead. \\ Security risks and dependence on validator distribution. } \\
  & SSLE & Employs advanced cryptographic techniques to ensure the selected validator's secrecy during leader election. & \parbox{6cm}{- In R\&D phase, no implementation is available.\\ - Increased complexity due to intricate cryptographic techniques. \\ - Shuffling and reorganizing of commitments might result in performance overhead. \\ - Increased latency. \\ - Scalability can be challenging.} \\
  & SnSLE & An alternative approach to SSLE that allows each validator (who shares commitments), a random opportunity to propose a block in every slot. & \parbox{6cm}{-  In R\&D phase, no implementation is available. \\ - Random selection based on commitments and hashes can be complex and errors can lead to undesirable and unpredictable outcomes. \\ - Similar to SSLE, the challenges like overhead, latency, and scalability will remain.  } \\
\bottomrule
\end{tabular}}\label{tab:solutions_comp}
\end{table*}

\paragraph{Secret Single Leader Election (SSLE)}
SSLE employs cryptographic techniques to ensure that only the selected validator knows its selection \cite{sle2023}. This is achieved through a process where each validator submits a commitment to a shared secret. The commitments are then shuffled and reorganized in a way that no one can map the commitments to specific validators while still allowing individual validators to identify their commitment. Subsequently, a commitment is randomly selected and the validators identify whether their commitment is selected or not, if a validator successfully identifies the chosen commitment, then they know about their turn to propose a block. One of the famous SSLE implementations is known as WHISK \cite{whisk}, which approaches the problem at the root level, where all modifications are made at the consensus layer while relaxing constraints such as latency. In this way, SSLE prevents DoS attacks by hiding the information about the validator selected for proposing a new block. 

\paragraph{Secret Non-single Leader Election (SnSLE)}
SnSLE is an alternative proposal for the secret leader election that provides each validator (sharing commitments) a random chance to propose a block in each slot. This random selection can simply be done through the RANDAO function, which is based on the idea that commitments (hashes) submitted by all validators are mixed to generate a sufficiently random number. This can also be used to select the next block proposer in SnSLE, e.g., the lowest-value hash. The range of valid hashes can be adjusted to infer the probability of individual validators being chosen within each slot. For instance, it imposes a constraint that the $\mathcal{}{h} < \frac{2^{256} \times 5}{N}$, where $N$ represents the total number of participating validators and $\mathcal{}{h}$ is the valid hash being generated. Therefore, the likelihood of each validator's selection will be $\frac{5}{N}$ for each slot, e.g., in this case, the chance of valid hash generation in each slot will be 99.3\% \cite{SNSLE}. It is worth noting that both SSLE and SnSLE are in the research phase and currently, there is no specified implementation of this proposal. However, SSLE and SnSLE are competing proposals that require more research, prototyping, and implementation. A comparison of different network and consensus layer techniques is provided in Table \ref{tab:solutions_comp}.

\subsection{Evaluating Privacy Preserving Routing Protocols}
In the literature, a few studies have evaluated different privacy-preserving P2P routing protocols using different quantified measures. For instance, to evaluate the anonymity guarantees of network-level privacy-preserving protocols, Serena et al. \cite{serena2021simulation} presented a simulator named LUNES-Temporal for the modeling of dynamic networks. They evaluated anonymity guarantees for two privacy-preserving routing protocols (Dandelion and Dandelion++). Their analysis reveals that these methods incur higher latency and less anonymity in delivering messages. Nonetheless, their simulator lacks modularity and does not support various Ethereum-specific intricacies that significantly alter the analysis of anonymity. To address this issue, an Ethereum network simulator named ethp2psim is presented in a recent study \cite{beres2023ethp2psim}, which can be used for the evaluation and deployment of P2P anonymity techniques. This simulator provides two capabilities: (1) it allows the implementation of new privacy-aware P2P routing techniques, and (2) it can be used for evaluating the privacy guarantees of anonymity techniques in routing messages.  

In addition to quantified measures, the use of probabilistic techniques to analyze the anonymity properties of communication protocols has been explored in the literature \cite{shmatikov2002probabilistic}. In \cite{danezis2009vida}, the authors presented the idea of using Bayesian inference for evaluating the anonymity of P2P routing protocols, they analyzed the anonymity of mixnets. Building upon their work, Sharma et al. \cite{sharma2022anonymity} presented a framework that employs Bayesian inference to derive probability distributions that establish connections between transactions and their potential originators. They defined transaction anonymity using calculated distributions, utilizing entropy as a metric to quantify the level of adversarial uncertainty regarding the identity of the originator. Furthermore, they performed experimental evaluations of Dandelion, Dandelion++, and lighting networks and demonstrated that none of these approaches provide the desired anonymity to users. They demonstrated that in a lightning network, even with just 1\% of strategically selected colluding nodes, the adversary can distinctly identify the originator for approximately 50\% of all transactions within the network. Whereas, in Dandelion, when an adversary manages to get control of 15\% of the nodes, the average level of uncertainty is limited to merely 8 potential originators. For the experimental evaluation, their network simulator is publicly accessible.\footnote{\url{https://github.com/pi-yush/anon-crypt}}

\begin{figure*}
    \centering
    \includegraphics[width=0.75\textwidth]{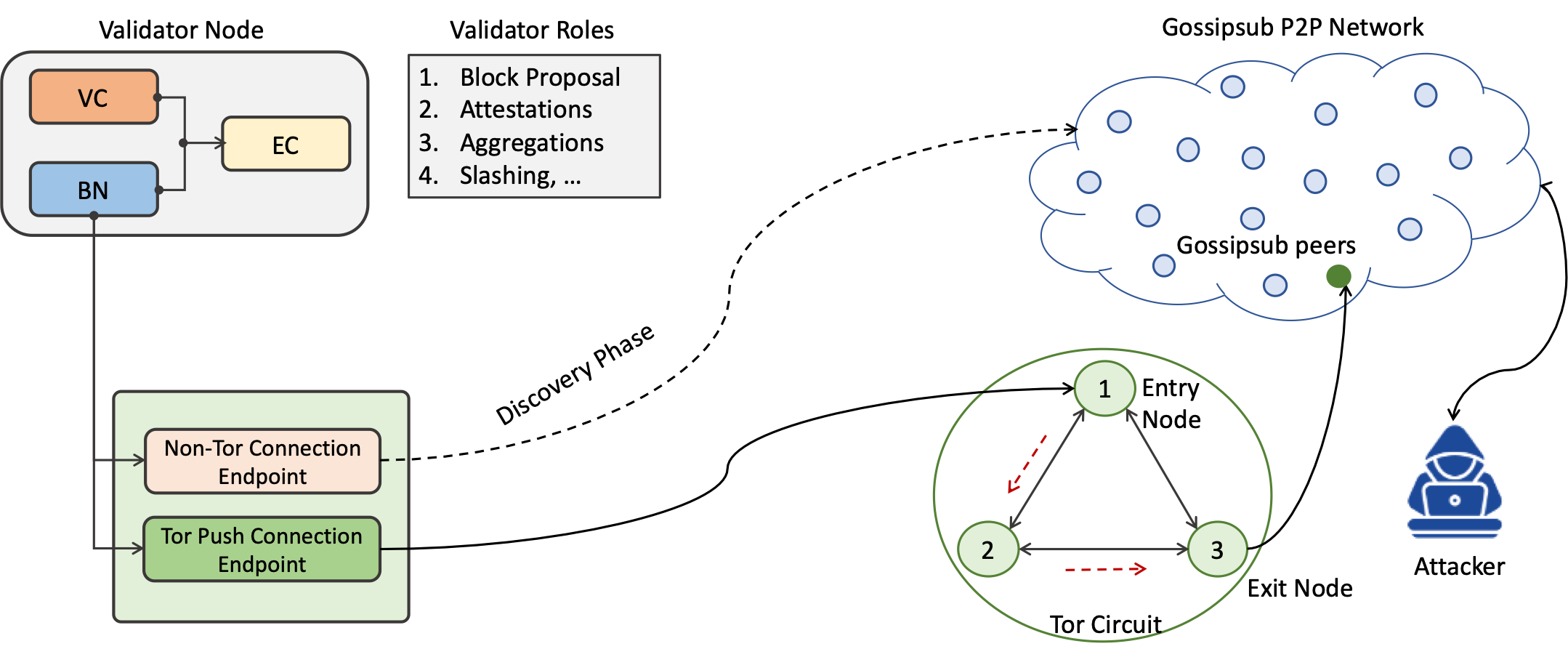}
    \caption{{Illustration of \emph{Tor push}  protocol that incorporates both non-Tor connection endpoint and \emph{Tor push}  connection endpoint. As shown in the figure, \emph{Tor push}  shifts all communication over the Tor network (except peer discovery).} }
    \label{fig:tor}
\end{figure*}

\subsection{{Analyzing Trade-off between Anonymity, Latency, and Bandwidth}}
In addition to the works focused on evaluating the overall performance of anonymity techniques, a few studies have attempted to quantify to which extent important parameters such as strong anonymity, low network bandwidth, and low network latency can be simultaneously achieved. In this regard, Das et al. \cite{das2018anonymity} performed a theoretical analysis of the fundamental constraints of privacy-preserving routing protocols and analyzed the relationship (i.e., the trade-off) between bandwidth, latency, and sender or recipient anonymity. Specifically, they derived necessary constraints for anonymity (upper bounds and lower bounds) and provided guidelines for improving existing anonymity techniques and improvising novel protocols. Their analysis suggests that only two of the three constraints can be simultaneously achieved.

\section{Validator Privacy via \emph{Tor push} }
\label{sec:proposed}

This section presents \emph{Tor push} \cite{vac46GOSSIPSUBTORPUSH}, a Tor-based message dissemination mechanism designed to mitigate validator deanonymization by obscuring the linkage between a validator’s network identifier (IP address) and its public validator's ID in the Ethereum peer-to-peer (P2P) network. \emph{Tor push}  follows a \emph{push-only} communication model that integrates with the existing GossipSub messaging layer while remaining fully backward compatible.

\subsection{Terminology}

We use the following terminology throughout this section.

\begin{itemize}
    \item \textit{GossipSub Messaging Network:} A publish-subscribe protocol is a communication protocol used in peer-to-peer (P2P) networks where nodes maintain a bounded number of peer connections to disseminate protocol messages.
    \item \textit{PubSub Topics:} Logical channels used to categorize and disseminate messages to interested peers.
    \item \textit{Beacon Node (BN):} An Ethereum client component responsible for consensus participation, message propagation, and network synchronization.
    \item \textit{Tor Connection Endpoint:} A network endpoint configured to transmit messages exclusively via the Tor network.
    \item \textit{Non-Tor Connection Endpoint:} A conventional network endpoint used for non-anonymized (non-Tor) communication.
\end{itemize}

\subsection{System Overview}

Figure~\ref{fig:tor} illustrates the high-level architecture of \emph{Tor push}. Each validator operates through its beacon node, which maintains two strictly segregated connection endpoints, namely a Tor connection endpoint and a non-Tor (conventional) connection endpoint. This separation is a fundamental design requirement and must be preserved at all times to prevent correlation attacks.

The non-Tor connection endpoint is used solely for peer discovery. It subscribes to relevant pub-sub topics and participates in the standard discovery mechanisms of the GossipSub network. Importantly, this endpoint does not originate or relay validator protocol messages and does not serve as a fallback channel for \emph{Tor push} traffic.

All validator-related protocol messages, including block proposals, attestations, aggregation publications, and synchronization messages, are transmitted exclusively via the Tor connection endpoint. By shifting validator-originated traffic onto Tor, \emph{Tor push} significantly increases the difficulty of linking a validator’s public identity to its underlying network location, while still allowing the validator to remain discoverable within the network.

\subsection{\emph{Tor push} Operation}
\emph{Tor push} is a push-only message dissemination mechanism that routes validator-originated traffic through the Tor anonymization network before it enters the GossipSub layer, thereby obscuring the link between a validator's public identity and its IP address. This subsection describes the operational details of \emph{Tor push} as deployed in our proof-of-concept implementation.

\emph{Tor push} messages use the same protocol identifier and wire format as standard GossipSub traffic, as specified in~\cite{vac46GOSSIPSUBTORPUSH}. 
Specifically, each message conforms to the standard libp2p pub-sub message structure, carrying fields for the sender identity, payload, sequence number, topic, and optional signature material. \emph{Tor push} does not introduce 
a dedicated protocol identifier; this design choice is deliberate, as it allows nodes that are oblivious to \emph{Tor push} to receive and 
process incoming messages according to the standard specification without any modification.

\emph{Tor push} operates in epochs, with a default epoch duration of 10 minutes. For each epoch, the validator proactively establishes a set of $D$ Tor circuits during the preceding epoch, where $D$ is set to the GossipSub out-degree parameter. Each circuit consists of three Tor relays, in accordance with the standard Tor protocol. Pre-establishing $D$ circuits reduces message latency at transmission time and improves robustness against circuit failures.

During an active epoch, the Tor connection endpoint uses these pre-established circuits to transmit messages via the Tor network using the SOCKS5 protocol. All \emph{Tor push} traffic is confined to a dedicated libp2p context, referred to as the \emph{Tor push} context, which must not share any data with the default context, including peer lists, and must not be reused for any other networking purpose; any such leakage would enable an observer to correlate Tor traffic with standard GossipSub traffic, thereby linking the validator's Tor connection to its public identity. Furthermore, only data messages are transmitted via \emph{Tor push}; control messages of any kind, such as GossipSub graft, must not be sent over \emph{Tor push} connections.

Upon receiving a \emph{Tor push} message, the receiving beacon node relays the message into the GossipSub network, where it is disseminated according to standard pub-sub rules. \emph{Tor push} is intentionally not implemented as a Tor onion service. Since the design goal is to hide network location rather than participation itself, avoiding onion services eliminates additional cryptographic overhead and limits message transmission to three Tor hops, thereby achieving a favorable balance between anonymity and performance. As a result, peers that do not explicitly support \emph{Tor push} can still receive, relay, and process \emph{Tor push} messages without modification, ensuring full backward compatibility with the existing Ethereum P2P stack.

\subsection{Security and Deployment Considerations}

The effectiveness of \emph{Tor push} relies critically on strict endpoint isolation. The Tor connection endpoint must never subscribe to pub sub topics or participate in peer discovery, and the non-Tor endpoint must never originate or relay validator protocol messages. Violating this separation risks enabling correlation between Tor and non-Tor traffic, undermining the anonymity guarantees of the system.

By decoupling peer discovery from message origination and enforcing push-only Tor-based transmission for validator messages, \emph{Tor push} preserves compatibility with existing Ethereum clients while significantly raising the cost of traffic analysis and targeted network layer attacks against validators. This design allows incremental deployment without requiring changes to the underlying consensus protocol or GossipSub message semantics.

\section{{Results and Discussions}}
\label{sec:results}
In this section, we present and discuss the implementation results of \emph{Tor push}.

\subsection{Notations} Let's assume the Tor connection endpoint is represented as $T_CE$ and the non-Tor connection endpoint is denoted as $N_CE$. The validator clients $VC_i$ are connected to both $T_CE$ and $N_CE$, where $VC_i$ uses $N_CE$ for communicating messages related to validator roles. Specifically, when we mention ``message proposal'', it encompasses tasks such as attestation, block proposal, and aggregations. In the network, we have $N$ number of nodes subscribed to a normal messaging network (non-Tor), and similarly, there are $M$ nodes that are running over the Tor network. Whenever a new block/attestation is proposed to the messaging network, $VC_i$ transmits this message to the Tor network via $T_CE$ using \emph{Tor push}. Then, a Tor circuit comprising three nodes is established to relay the message received from the $VC_i$ to the destination node. These three nodes are randomly selected from the $M$ nodes that are running over the Tor network. The first node that receives the message from $VC_i$ is known as the entry node $E_T$ and the last node that relays the message to the messaging network node is known as the exit node $E_X$. The anonymity set is represented as $A$, which is defined as a set of all possible source nodes of a particular communication, indistinguishable from each other to the observer that could identify any specific entity as the actual originator.

\subsection{Experimental Setup and Implementation Details}
We used a Linux testbed machine to run a validator attached to a Nimbus beacon node that integrates with Tor to broadcast attestations and blocks. Nimbus is a secure lightweight client for the Ethereum network with efficient resource consumption. The Nimbus beacon node connects to the beacon chain network and syncs historical data. The validator is attached to the beacon node to perform its activities like signing attestation and producing blocks. The beacon node establishes Tor connections with discovered peers and makes these available beforehand for broadcasting any attestation or block over Tor. This adds additional latency due to message routing over Tor-circuits before it reaches the destination peer. We set up both the sending beacon node (using Tor) and receiving beacon node (without Tor) on the same testbed machine to measure the minimal latency possible. The study was conducted in 2024, based on a deployment maintained from 5 October to 21 October 2023.

\begin{figure*}[!ht]
    \centering
    \subfigure[Non-Tor Single Validator]{\includegraphics[width=0.45\textwidth]{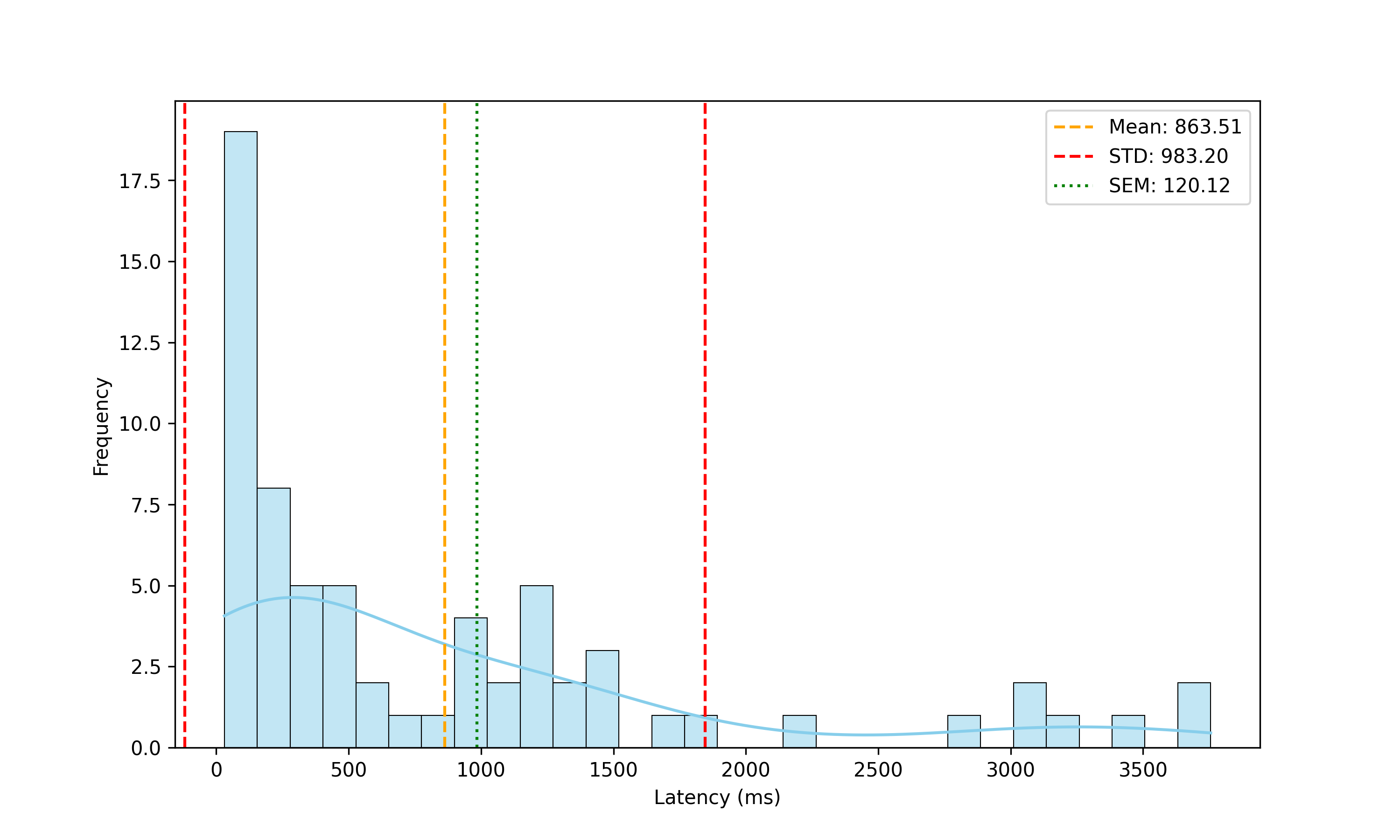}}
    \subfigure[Non-Tor Ten Validators]{\includegraphics[width=0.45\textwidth]{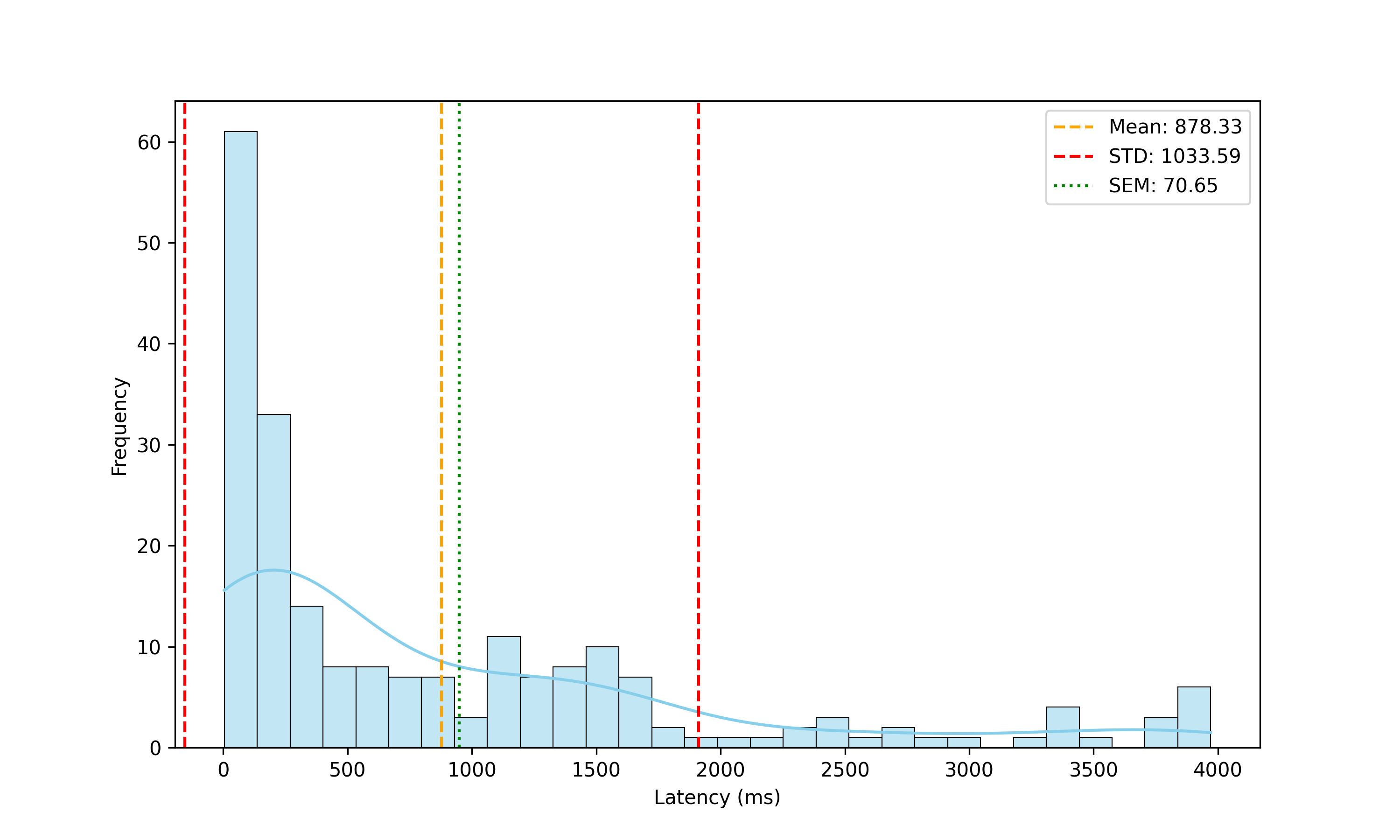}}
    \subfigure[Tor Single Validator]{\includegraphics[width=0.45\textwidth]{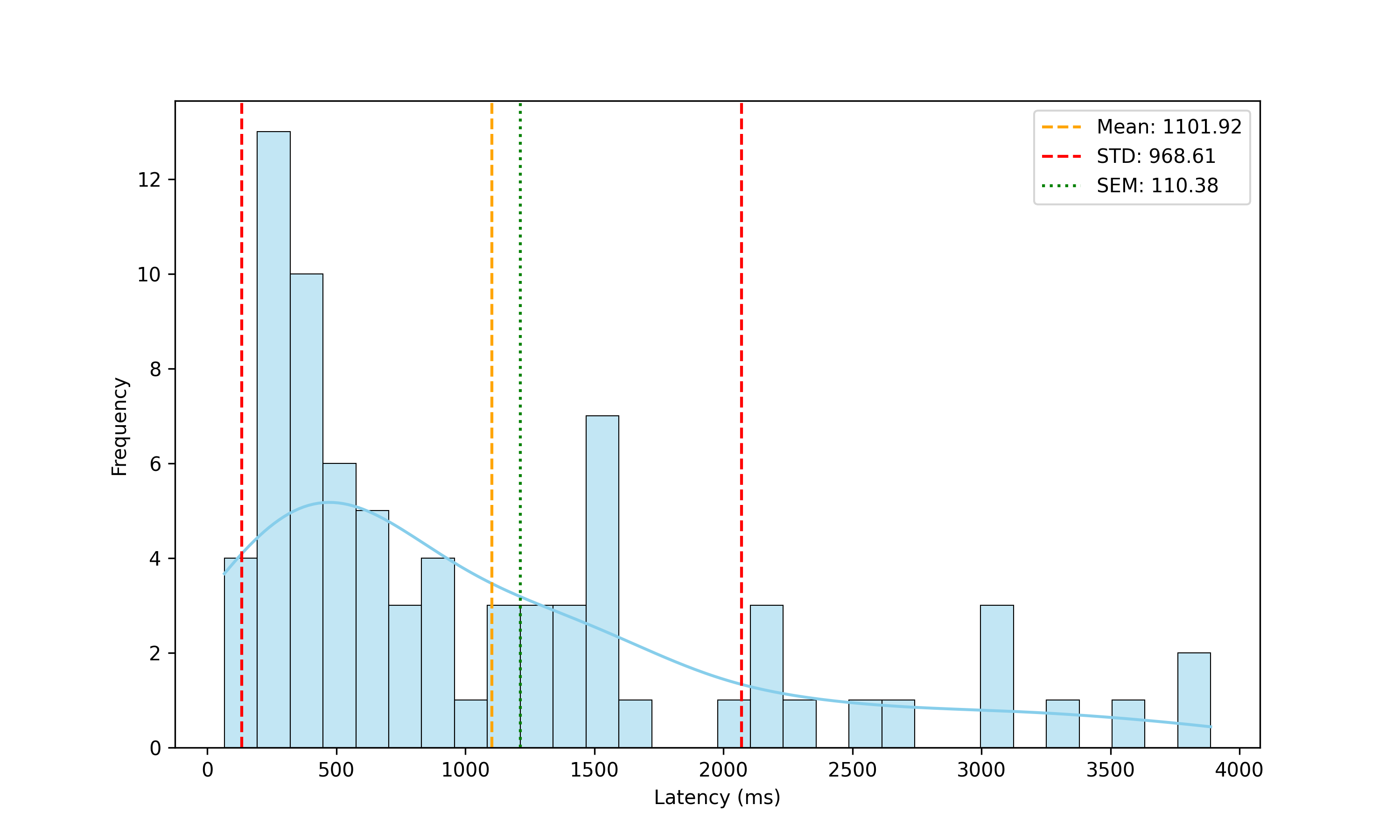}}
    \subfigure[Tor Ten Validators]{\includegraphics[width=0.45\textwidth]{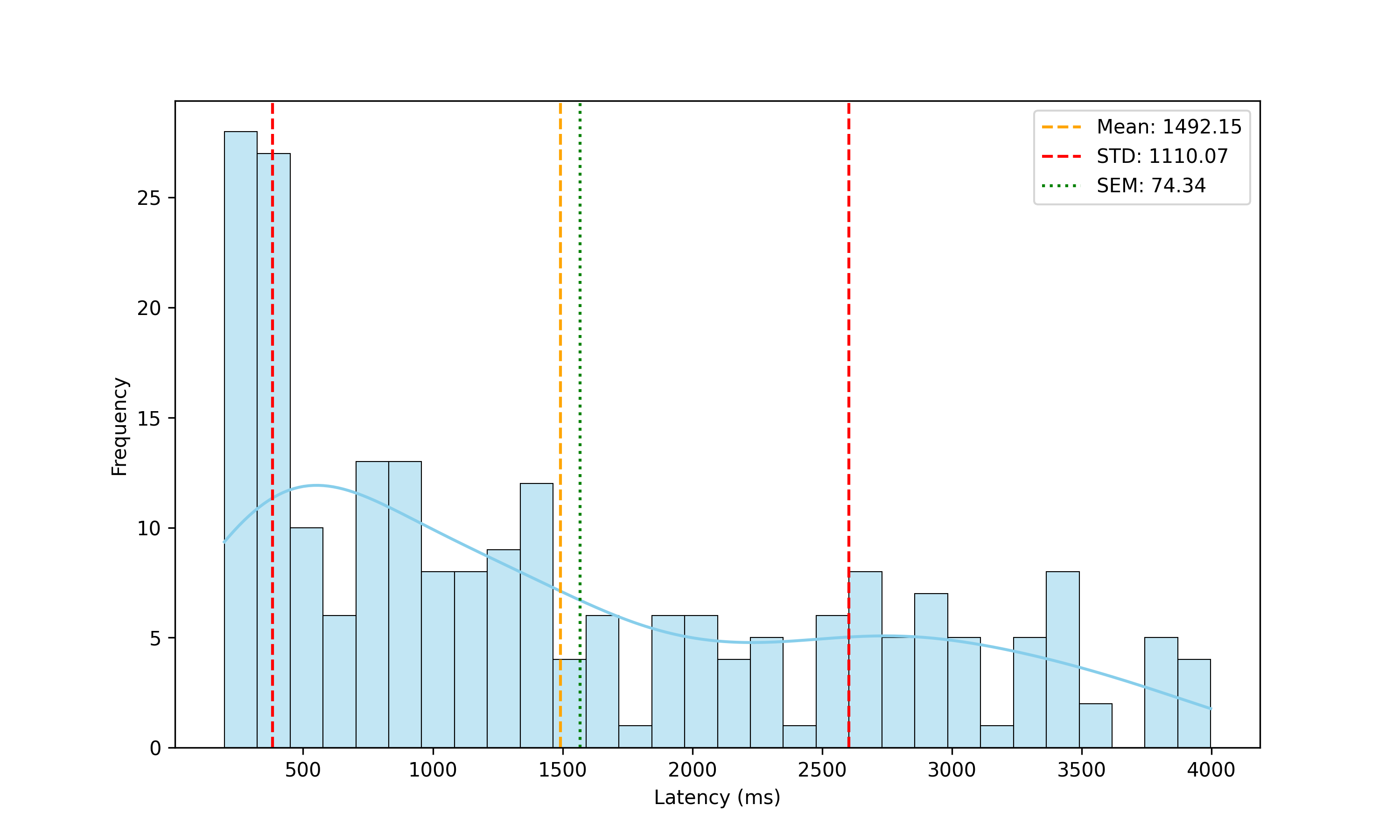}}
    \caption{{Latency analysis of non-Tor (Direct) and Tor message delivery (i.e., time to reach from sender to receiver for each attestation) in Ethereum blockchain (using single validator and ten validators). The x-axis represents time in milliseconds and the y-axis denotes the number of attestations. Legend: Mean: Average latency; STD: Standard Deviation; and SEM: Standard Error of Mean.}}
    \label{fig:latency}
\end{figure*}

\subsection{{Performance Analysis of \emph{Tor push} }}

\subsubsection{{Performance Metrics}}
In the literature, various metrics have been used to evaluate the effect of incorporating privacy protection techniques into blockchain networks, e.g., delay, latency, and effectiveness. To quantitatively evaluate the efficacy of \emph{Tor push} in performing attestations in the Ethereum blockchain, we have used the following performance metrics. The evaluation covers attestations only; block proposal performance is deferred to future work.

\paragraph{{Latency}}
We measured the latency (i.e., time taken) by the messages (i.e., attestations) from the sender to the receiver using the following formulation. Let's assume the latency experienced by a validator when communicating directly with the blockchain network without Tor is $L_{base}$. The additional latency introduced by each Tor node for relaying the message is given as $L_{hop}$, where the number of hops (i.e., Tor nodes between the validator and the blockchain network) is $N_{hops}$. The size of the message (i.e., attestations) being transmitted through Tor is $S_{msg}$. The total latency ($L_{Tor}$) experienced by the validator when using Tor for making attestations to the blockchain using a network having a bandwidth of $C_{bw}$ will be calculated as: 

\begin{equation}
    L_{Tor} = L_{base} + (N_{hops} \times L_{hop}) + \frac{S_{msg}}{ C_{bw}}
\end{equation}

In the above equation, the term ($\frac{S_{msg}}{ C_{bw}}$) takes into account the impact of message size on latency. It represents the time taken to transmit the message ($S_{msg}$) over the available bandwidth ($C_{bw}$). This accounts for the additional time required to send larger messages through the Tor network, which can contribute to increased latency.

\begin{figure}[!ht]
    \centering
    \includegraphics[width=0.483\textwidth]{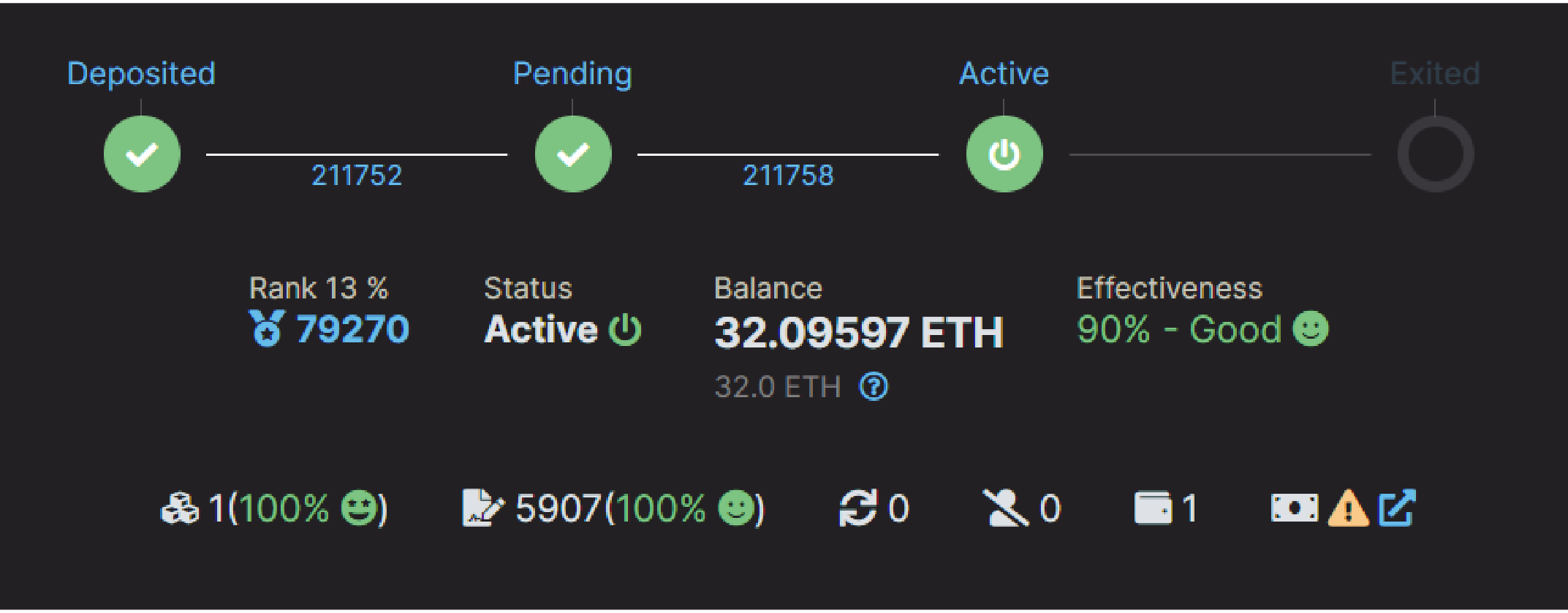}
    \caption{{{Effectiveness of a single Tor validator in the deployment.}}}
    \label{fig:effectiveness}
\end{figure}

\paragraph{{Effectiveness}}
Attestation Effectiveness is calculated as the ratio of the difference between the earliest(minimum) possible slot and the actual inclusion slot from the attestation slot. An effectiveness score of 1 indicates immediate inclusion and optimal validator reward, while a lower score denotes delayed inclusion and reduced reward.
    \begin{equation}
    {E =\frac{S_{\text{eis}} - S_{\text{as}}}{S_{\text{ais}} - S_{\text{as}}}} 
    \end{equation}
    {Where, $E$ is the effectiveness, ($S_{\text{ais}}$) refers to the minimum possible slot in which a validator’s attestation can be included in the blockchain. ($S_{\text{as}}$) is the slot in which the attestation is due or is made by the validator. ($S_{\text{ais}}$) is the slot in which the attestation is included in the blockchain. Fig. \ref{fig:effectiveness} shows the effectiveness of a single Tor validator with an effectiveness of 90\%} \footnote{Note that we only maintained the deployment from 5 October 2023 to 21 October 2023 during the evaluation of \emph{Tor push}.}

\paragraph{{Attestation Miss Frequency}}
Attestation Miss frequency is measured as the number of times a validator fails to perform an attestation that is not included.

\subsubsection{{Empirical Evaluation}}
\begin{figure}[!ht]
    \centering
    \includegraphics[width=0.45\textwidth]{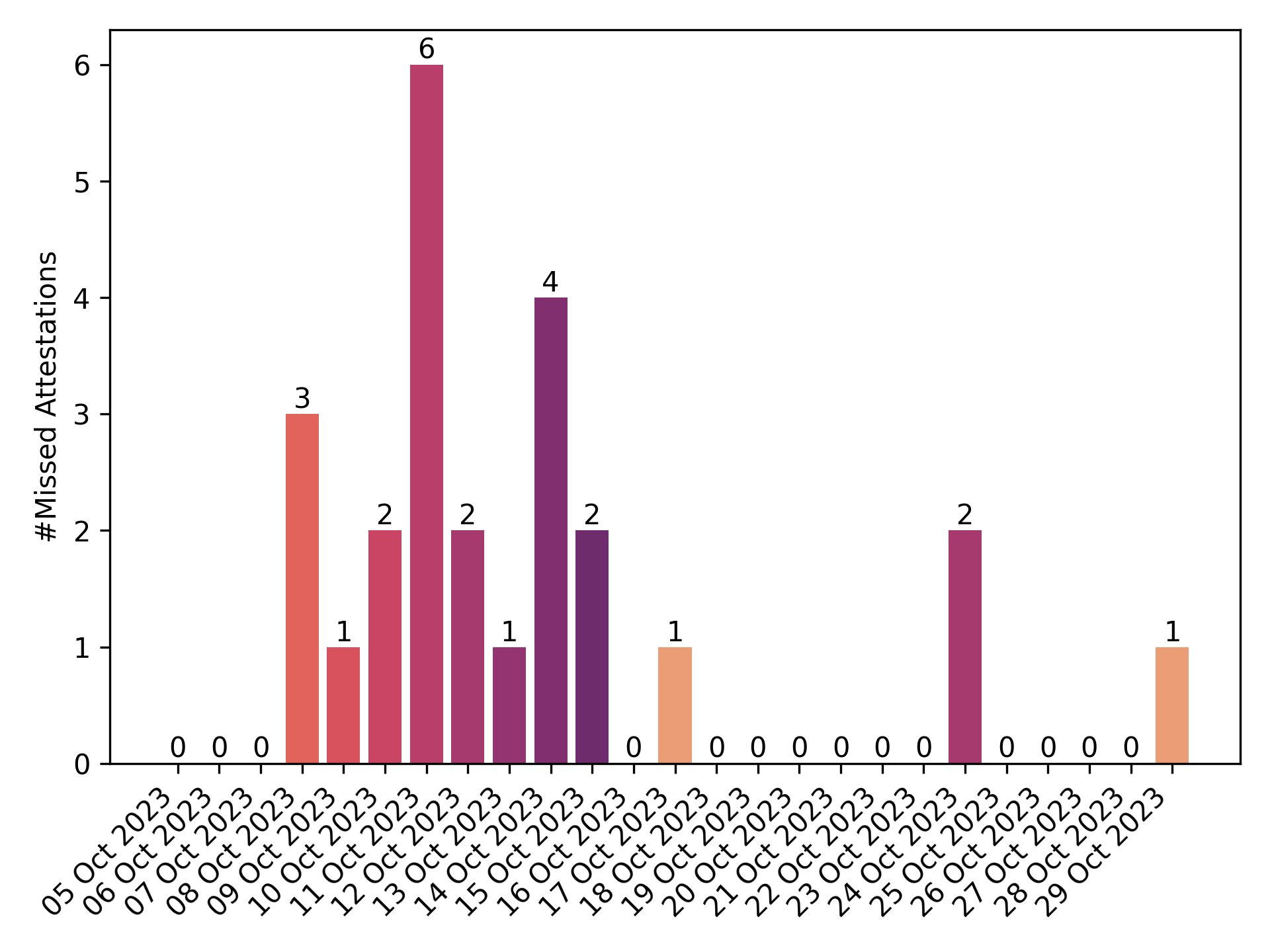}
    \caption{{Demonstrating the number of missed attestations daily for some time when \emph{Tor push} is integrated into the Ethereum blockchain network.}}
    \label{fig:missed}
\end{figure}

\begin{figure}[!ht]
    \centering
    \includegraphics[width=0.45\textwidth]{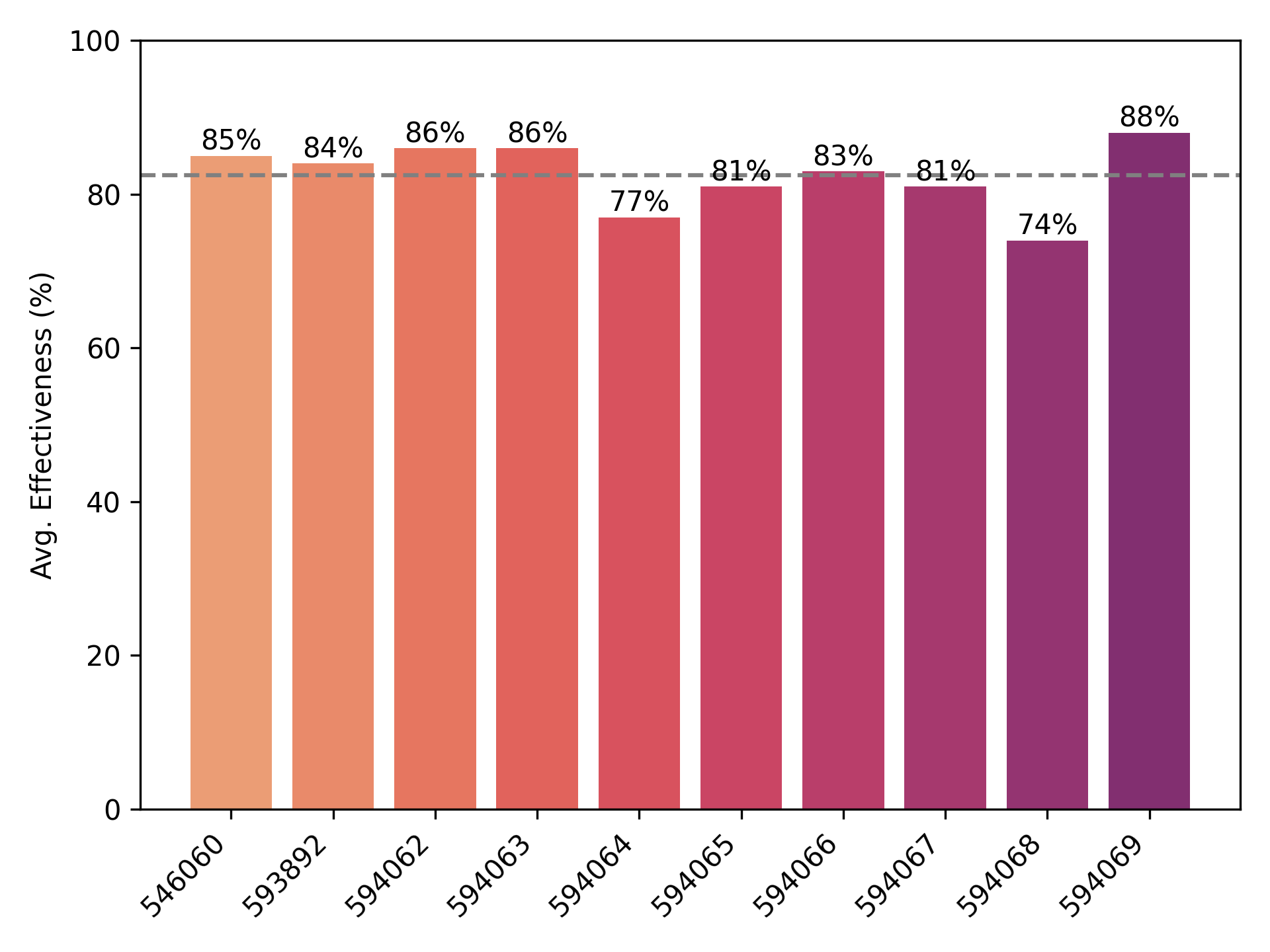}
    \caption{{Comparative analysis of ten Tor-based validators in terms of average effectiveness, where the average of all validators is 82.5\%.}}
    \label{fig:effect}
\end{figure}

\begin{figure}[!ht]
    \centering
    \includegraphics[width=0.5\textwidth]{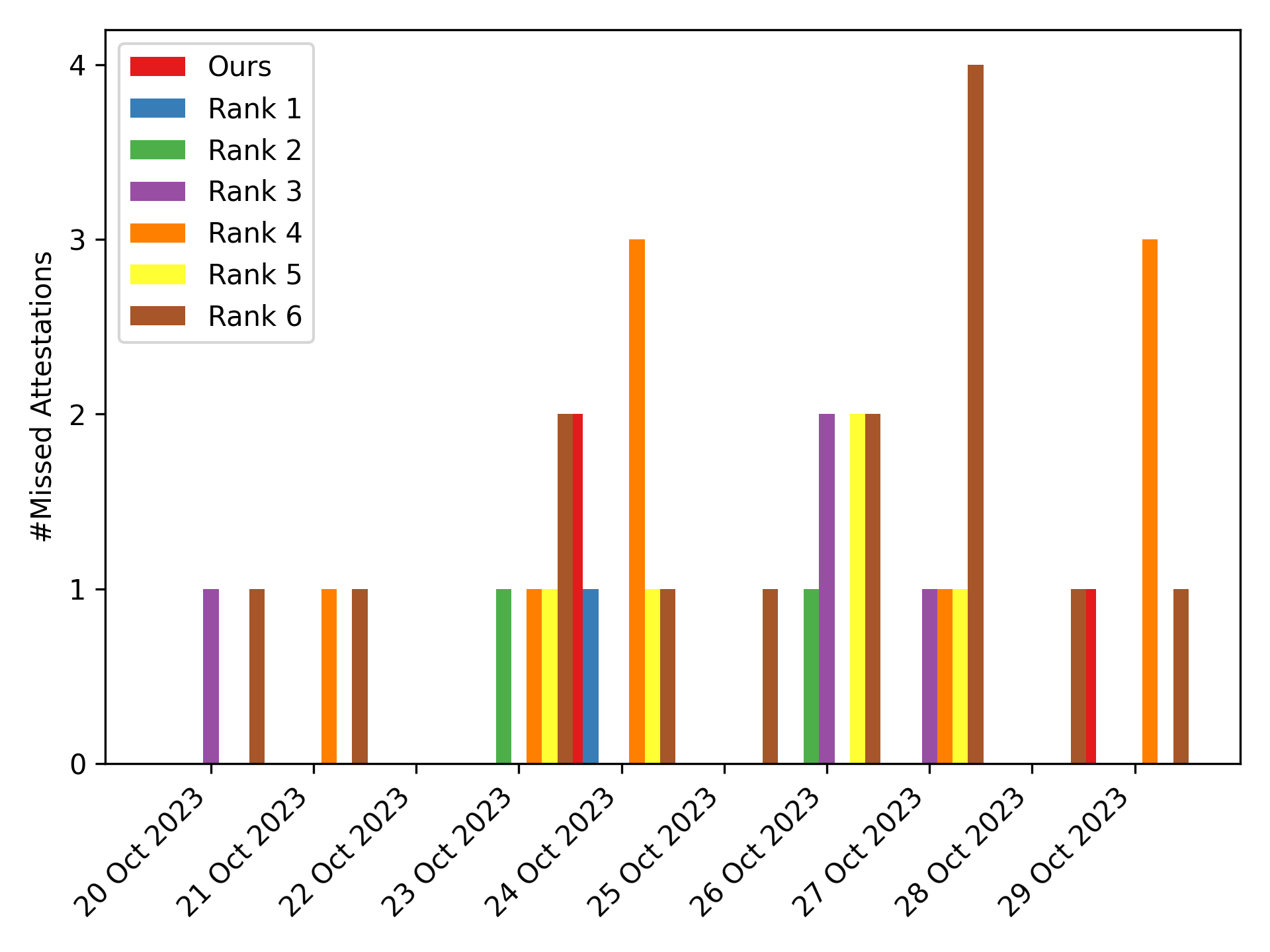}
    \caption{{Comparison of our Tor-based validator with top-6 (non-Tor) validators on the leaderboard in terms of number of missed attestations over 10 days.}}
    \label{fig:missed_comp}
\end{figure}

Fig. \ref{fig:latency} (a) demonstrates the histogram for latency in ms (x-axis) vs. the number of attestations (y-axis), where the latency is measured in milliseconds for a single validator. On average \emph{Tor push} results in an overhead of approximately 238.41 ms, which is not significant. The average \emph{Tor push} latency is 1101.92 ms as compared to the average latency of direct communication which is 863.51 ms for 78 attestations having different sizes. To evaluate the scalability of \emph{Tor push}, we ran ten \emph{Tor push} validators on a single beacon node. The results for this analysis are shown in Fig. \ref{fig:latency} (c) and (d), which depict average latency for attestations made using normal and Tor networks. Note that the average latency for Tor-based attestation is 1492.15 ms and for the average latency for the normal network is 878.33 ms, where Tor added an overhead latency of 613.82, which is not significant. 
In addition to the latency, we also quantified the number of missed attestations daily, which is shown in Fig. \ref{fig:missed}. Note that during the period of 5 October 2023 to 21 October 2023, a total of 9548 have been successfully registered, while only a small fraction, i.e., 0.23\% attestations were missed over this period. {In addition, we have measured the average effectiveness of ten validators in Fig. \ref{fig:effect}, where the average effectiveness of ten validators is 82.5\%.} We also perform a performance comparison of our Tor-based validator with top-6 non-Tor validators from the Nimbus leaderboard, which is shown in Fig. \ref{fig:missed_comp}. It is evident from the figure that our Tor-based validator is performing almost comparable to the top validators. Top validators that are not running over Tor are also missing attestations (see Fig. \ref{fig:missed_comp}). 


\subsection{{Security Analysis}}
Given that \emph{Tor push} relies on the Tor network, it inherits the inherent vulnerabilities associated with Tor into our system. In this section, we analyse possible security attacks on our solution independently and/or in conjunction with vulnerabilities associated with Tor (a summary is provided in Table \ref{tab:security_summary}).

\begin{table*}[]
\centering
\caption{A summary of security attacks applicable to \emph{Tor push}, their relevance to the system, and mitigation strategies. Mitigations marked as future work are not yet implemented.}
\label{tab:security_summary}
\scalebox{0.98}{
\begin{tabular}{p{3.5cm}p{6.6cm}p{6cm}} 
\toprule \multicolumn{1}{c}{\textbf{Attack}} &
  \multicolumn{1}{c}{\textbf{Applicability to Tor push}} &
  \multicolumn{1}{c}{\textbf{Mitigation Strategy}} \\ \midrule
Guard-Only Deanonymization &
  An attacker controlling a significant fraction of Tor guard bandwidth could compromise the unlinkability provided by \emph{Tor push}, without requiring exit node control. &
  Guard diversity — a more diverse Tor relay pool reduces the guard bandwidth fraction any single adversary can control. \\ \midrule
Targeted Guard Disruption &
  Disrupting a validator's guard just before its assigned slot guarantees a missed proposal. &
  Distribute circuits across distinct guards. \\ \midrule
DoS Attack on Tor &
  Ethereum's public slot schedule amplifies targeted DoS — an adversary needs only act during a validator's assigned 12-second slot. &
  Building circuits across diverse exits provides resilience; the attacker must disrupt all these exits simultaneously. \\ \midrule
End-to-End Correlation &
  Ethereum's public slot schedule makes correlation more tractable — timing windows are known in advance, reducing statistical uncertainty. &
  Cover traffic; remains future work. \\ \midrule
Fingerprinting Attacks &
  Ethereum's protocol rhythm — attestations at regular intervals, proposals at scheduled slots — creates a recognizable fingerprint. &
  Padding and dummy messages; remains future work. \\ \midrule
Low-Resource Routing Attack &
  Adversarial nodes advertise false high bandwidth; a successful selection into a circuit exposes all validator activity for the full epoch. &
  Tor's bandwidth verification partially mitigates; Ethereum-specific path selection remains future work. \\ \midrule
Side-Channel Attacks &
  The public slot schedule enables passive latency-based geolocation over several epochs. &
  No mitigation identified. \\ \midrule
Reputation-Based DoS on Receiving Beacon Nodes & Adversaries inject spam via receiving beacon nodes, causing honest peers to blacklist them; creates a fate-sharing condition.
   & Modify peer score protocol; further examination required. \\ \midrule
Peer Discovery & Discovery runs outside Tor, exposing the validator's real IP; acts as a self-identification signal.
   & Run peer discovery over Tor; remains future work. \\ \midrule
Discovery--Connect Anomaly & A malicious discovery node assigns unique ports per requester IP; deterministically links the Tor connection to the requester. & Run peer discovery over Tor; remains future work. \\ \midrule
Exploiting Roll-out Phase & Peer discovery leaks in which validators use \emph{Tor push}; during rollout the small pool of known users collapses the anonymity set. & Run peer discovery over Tor; broader network adoption. \\ \midrule
Exploiting Peer Scores & Malicious actors lower the peer scores of legitimate receiving beacon nodes, leading to their blacklist, blocking validator message propagation. & Modify peer score protocol; further examination required. \\ \bottomrule
  
\end{tabular}}
\end{table*}

\subsubsection{Guard-Only Deanonymization}
\label{sec:guard-only-deanon}
In generic anonymous networks, deanonymization requires controlling both entry and exit nodes, giving a compromise probability of $f^2$ for an adversary controlling fraction $f$ of the network \cite{alsabah2016performance}. In the Ethereum context, however, exit node control is unnecessary: block proposals are broadcast over the public gossip network, so any observer can confirm a validator's action without observing the exits. Specifically, when a validator's circuit passes through a malicious guard, the guard observes the connecting IP address and correlates traffic bursts from that address with the publicly known slot schedule, revealing the validator's identity. The gossip network further confirms the corresponding block proposal, providing ground truth at no additional cost. Together, these observations allow an adversary controlling a fraction $g$ of guard bandwidth to deanonymize a validator with probability $g$, without requiring exit node control. Since Tor clients (\textit{i.e.,} validators using \emph{Tor push}) retain guards for a few months at least, a successful compromise results in persistent exposure for that duration — no circuit rotation strategy can mitigate this. Guard diversity is therefore the most effective countermeasure: a larger and more diverse Tor relay pool reduces the guard bandwidth fraction any single adversary can control, making broad Tor network health the primary structural defense against this class of attack. 

\subsubsection{Targeted Guard Disruption}
Interrupting the service of a Tor guard may temporarily block validators associated with that guard, forcing a circuit rebuild. Re-establishing connectivity can take from tens of seconds to several minutes; Tor clients may wait up to the default 60-second circuit build timeout before abandoning a failing path \cite{tor-path-spec}, far exceeding Ethereum's 12-second slot window. In the context of \emph{Tor push}, this creates a targeted liveness risk: an adversary who has identified a validator's guard node via the mechanism described in Section \ref{sec:guard-only-deanon} can perform a targeted DoS against that guard in the seconds before the validator's assigned slot, guaranteeing a missed proposal at minimal cost. Pre-building $D$ circuits does not help since multiple circuits share the same primary guard — a single targeted disruption can cause total liveness failure for that validator. A potential mitigation would be to explicitly configure \emph{Tor push} to distribute circuits across distinct guards; however, this is not default Tor behavior and would require protocol-level enforcement.


\subsubsection{{DoS Attack on Tor}}
Untargeted DoS attacks on \emph{Tor push} require significant resources — \emph{Tor push} employs multiple circuits and the Tor network is highly robust, meaning an adversary would need to control a substantial fraction of bandwidth to cause meaningful disruption. Targeted DoS attacks, however, are amplified by Ethereum's public slot schedule: an adversary need only act during a validator's assigned 12-second slot rather than sustaining prolonged disruption. On the exit side, building circuits across diverse exits provides resilience since the attacker must disrupt all $D$ exits simultaneously. The guard-side variant, discussed in Section \ref{sec:guard-only-deanon}, represents a more critical exposure and is addressed there.

\subsubsection{End-to-End Correlation}
In generic Tor, end-to-end correlation requires observing traffic at both guards and exit nodes using statistical techniques such as packet counting \cite{serjantov2003passive}, moving window averages \cite{levine2004timing}, or powerful correlation methods \cite{murdoch2007sampled}. The scheduled nature of Ethereum validator activity makes this correlation deterministic rather than statistical, as the public slot schedule gives precise timing windows to check for correlated traffic. Cover traffic is a potential mitigation but remains future work in the current implementation.

\begin{table*}[!ht]
\centering
\caption{Comparison of this techreport with similar existing work (analysis of Bitcoin blockchain security over Tor network).}
\label{tab:my-table}
\scalebox{0.9}{
\begin{tabular}{p{1.5cm}p{5cm}p{1.5cm}p{4cm}p{3cm}p{3cm}}
\toprule
\multicolumn{1}{c}{\textbf{Ref.}} & \multicolumn{1}{c}{\textbf{Description}} & \multicolumn{1}{c}{\textbf{Blockchain}} & \multicolumn{1}{c}{\textbf{Attack Surface(s)}} & \multicolumn{1}{c}{\textbf{Attack(s)}} & \multicolumn{1}{c}{\textbf{Solution(s)}} \\
\midrule
Biryukov et al. \cite{biryukov2015bitcoin} & \multicolumn{1}{c}{\parbox{3cm}{Evaluated Bitcoin's security over the Tor network.}} & Bitcoin &  \multicolumn{1}{c}{\parbox{4cm}{Bitcoin and Tor-specific vulnerabilities are leveraged to compromise Tor anonymity.}} & \parbox{3cm}{- Delinking users using fingerprinting via message cookies. \\ - Leveraging Bitcoin's anti-DoS protection to block Tor exit nodes.} & \multicolumn{1}{c}{N/A} \\ \midrule

Kelen et al. \cite{kelen2023integrated} & \multicolumn{1}{c}{\parbox{3cm}{Studied the feasibility of integrating onion routing to maintain Ethereum's validator privacy.}} & Ethereum &  \multicolumn{1}{c}{\parbox{4cm}{Only provided the theoretical analysis of validator privacy and analyzed the applicability of using onion routing.}} & \parbox{3cm}{- First hop detection. \\ - Correlation attack. \\ - Spam protection. \\} & Provided theoretical guarantees for validator privacy protection. \\ \midrule

This techreport & \multicolumn{1}{c}{\parbox{3cm}{ \textit{Tor push} and provides an empirical evaluation in terms of latency and effectiveness. Also, theoretically analysed the potential attacks and discussed their mitigation strategies.}} & Ethereum & \multicolumn{1}{c}{\parbox{4cm}{Vulnerable to Tor-specific attacks (as Tor vulnerabilities are inherited directly).}} & \parbox{3cm}{- Ethereum's peer score can be used to restrict certain nodes. \\ - Attacks exploiting Tor's vulnerabilities.} & \parbox{3cm}{- Modify peer score protocol; further examination required. \\ - Use custom routing to reach Tor and source-level variations.} \\
\bottomrule
\end{tabular}}\label{tab:sota}
\end{table*}

\subsubsection{{Fingerprinting Attacks}}
Protocols that contain unique patterns are more susceptible to fingerprinting attacks when utilized over the Tor network. Both malicious guards and exits have the potential to identify these distinctive patterns at the end points, potentially linking the sender to the receiver. Ethereum validator traffic is particularly susceptible, as its protocol-defined rhythm — attestations at regular intervals and proposals at scheduled slots — provides a recognizable fingerprint observable within a single circuit. Biryukov et al. \cite{biryukov2015bitcoin} demonstrated that even low-resource attackers can assess communication patterns over Tor and link them to users. Since circuits are refreshed per epoch rather than per transaction, all messages within a 10-minute window share the same circuit — providing an extended observation window. Padding techniques and dummy messages are the appropriate mitigations but remain future work.

\subsubsection{Low-Resource Routing Attack}
Low-resource adversarial nodes can advertise false high bandwidth to increase their probability of selection in Tor circuits — a low-resource routing attack \cite{bauer2007low}. Tor partially mitigates this through bandwidth verification by directory authorities — relay bandwidth is measured independently rather than self-reported, limiting the effectiveness of false advertising. However, this protection is imperfect \cite{jansen2021accuracy}, and a successfully selected malicious node can observe circuit traffic. In \emph{Tor push}, since circuits are reused for the entire 10-minute epoch, a single successful selection at circuit construction time provides an observation window for all validator activity within that epoch. Designing path selection constraints specific to Ethereum validators could further reduce this risk. 

\subsubsection{Side-Channel Attacks}
Several side-channel attacks on Tor have been studied in the literature — (1) throughput fingerprinting attacks that exploit correlations between circuits sharing the same bottleneck node \cite{mittal2011stealthy}, (2) congestion attacks that use modulated traffic patterns to identify requesting peers \cite{murdoch2005low}, and (3) latency-based attacks, which infer circuit relationships by comparing latency distributions \cite{hopper2010much}. Both throughput fingerprinting and congestion attacks are costly in practice, requiring sustained active analysis, and are no different in the Ethereum context. However, for latency-based attacks, the Ethereum public slot schedule enables passive measurement without active probing — a malicious guard receiving traffic at a consistent offset after slot boundaries can estimate the validator's network distance \cite{hopper2010much}, narrowing the anonymity set to a geographic region over several epochs.

\subsubsection{{Reputation-Based DoS on Receiving Beacon Nodes}}
Adversaries can inject spam into the Ethereum gossip network via receiving beacon nodes, causing honest peers to lower the IP-based reputation of those nodes within the GossipSub scoring framework and eventually blacklist them — a reputation-based DoS \cite{vyzovitis2020gossipsub}. Since multiple validators may forward messages to the same beacon node, this creates a fate-sharing condition: a single attacker targeting one beacon node can simultaneously block all validators using it from propagating messages into the gossip network. Simply blocking high-transmission peers is inadequate, as legitimate \emph{Tor push} nodes share these beacon nodes. To mitigate this, we advocate a modified peer score protocol where nodes are throttled rather than completely blocked; however, further examination is required for an appropriate solution.


\subsubsection{{Peer Discovery}}
\label{sec:peer-discovery}
In the current implementation, peer discovery runs outside Tor, exposing the validator's real IP address during the discovery phase. Furthermore, querying specifically for \emph{Tor push} capable peers acts as a self-identification signal — an adversary operating a bootnode or crawler can build a target list of validators with anonymity-seeking intent. Running peer discovery over Tor is the appropriate mitigation and is planned as future work.

\subsubsection{{Discovery--Connect Anomaly}}
A modified discovery node can assign a unique port to each requester and store the mapping internally. When the same requester later connects via Tor to that assigned port, the attacker deterministically links the Tor connection to the original IP address. The table below illustrates this mapping.

\begin{table}[H]
    \centering
    \begin{tabular}{cc}
        \toprule
        \textbf{Request from} & \textbf{Response} \\ \midrule
        \texttt{22.11.141.44} & \texttt{X.X.X.X:4000} \\
        \texttt{13.12.35.101} & \texttt{X.X.X.X:4001} \\
        \texttt{121.25.11.102} & \texttt{X.X.X.X:4002} \\
        \texttt{44.11.13.6} & \texttt{X.X.X.X:4003} \\ \bottomrule
        \end{tabular}
\end{table}

Running peer discovery over Tor is the appropriate mitigation for both this attack and the peer discovery vulnerability discussed in Section \ref{sec:peer-discovery}; however, this remains an open gap in the current implementation and is planned as future work. 

\subsubsection{Exploiting Roll-out Phase}
During the initial deployment of \emph{Tor push}, the anonymity set is small. As noted in Section~\ref{sec:peer-discovery}, peer discovery leaks which validators are using \emph{Tor push}. During rollout, when only a handful of validators use \emph{Tor push}, this discovery-level exposure collapses the anonymity set — the pool of plausible message originators reduces to a few known validators. This limitation is inherent to the rollout phase and is compounded by the unresolved discovery vulnerability; both are addressed by running peer discovery over Tor. Note that the proof-of-concept deployment described in Section~\ref{sec:results} operated in this roll-out phase, with a small number of \emph{Tor push} validators and no broader adoption on the network at the time, resulting in a minimal anonymity set.

\subsubsection{Exploiting Peer Scores}
\label{sec:exploiting-peer-scores}
Peer scores in the Ethereum gossip network serve as a reputation mechanism — nodes use them to select trustworthy neighbors for message relay \cite{vyzovitis2020gossipsub}. Malicious actors can exploit this by artificially lowering the peer scores of legitimate receiving beacon nodes, causing honest peers to blacklist them and preventing validator messages from propagating. Disabling peer scores mitigates this exploitation; however, it removes the primary defense against gossip spam and Sybil attacks, since any node can then send arbitrary traffic without reputation consequences. Therefore, a replacement admission mechanism — such as rate limiting — would be required to maintain network integrity while eliminating the peer score exploit vector.

\subsection{Comparison with Existing Similar Works}
In this section, we perform a comparison of this techreport with existing similar studies that aim to utilize Tor network to achieve validators' anonymity. Kelen et al. \cite{kelen2023integrated} theoretically analyzed the feasibility of integrating onion routing to maintain Ethereum's validator privacy. They provided theoretical guarantees for validator privacy protection by considering three attacks, first hop detection, correlation attack, and spam protection. Although their work is very similar to our techreport, however, it notably lacks experimental analysis of Ethereum's operation over the Tor network. In a similar study, Biryukov et al. \cite{biryukov2015bitcoin} analyzed the anonymity provided by Tor in Bitcoin P2P transactions. Their study identified vulnerabilities inherent in both Bitcoin and Tor protocols, revealing potential compromises in Tor's anonymity. Their findings suggest that even adversaries with limited resources can potentially monitor the traffic flows between nodes operating Bitcoin over Tor. Consequently, the authors suggested not to integrate Tor with the Bitcoin blockchain network, as it may not provide the desired level of anonymity and security for Bitcoin validators. Similarly, certain features of the Ethereum network, akin to those exploited in Bitcoin, could be leveraged to realize DoS on the Ethereum network running over Tor. For instance, Ethereum's peer score mechanism can be exploited by malicious actors to restrict certain nodes from receiving messages. To mitigate this risk, we suggest modifying the peer score protocol when integrating Tor with the Ethereum network (see Section~\ref{sec:exploiting-peer-scores}). 




\section{Conclusion}
\label{sec:concs}

In this techreport, we analysed \emph{Tor push} focusing on improving validators' location privacy, especially with respect to IP unlinkability in the Ethereum blockchain using Tor.
We performed a review of existing network layer techniques and analyzed \emph{Tor push}, a method for mitigating linking between the public key of validators and their IP address.
We presented a working, deployed proof-of-concept implementation of performing attestations on an active test Ethereum network and analyzed the performance in terms of latency and effectiveness. To evaluate \emph{Tor push}, we executed ten Ethereum validators on a single beacon node and we observed an average latency overhead of approximately half a second with an average effectiveness of around 82\%. We believe this improves location privacy, especially IP unlinkability without adding significant latency on the network. This techreport is a first step towards getting more insights on the feasibility of Tor-based solutions for Validator privacy. In our future work regarding \emph{Tor push}, we plan to improve the latency evaluation to incorporate block proposals and aggregations across various geographic locations. In addition, to mitigate the risk of discovery connect anomaly, we plan to shift the discovery phase over the Tor network in our future implementation. We also plan to quantify the anonymity guarantees of \emph{Tor push} by realizing actual attacks on it.

\section*{Acknowledgement}
We would like to thank Muhammad Umar Farooq for his valuable feedback that contributed to the improvement of this work.

\bibliographystyle{IEEEtran}
\bibliography{sample}

\end{document}